\documentclass [reqno,10pt]{amsart}
\usepackage{amsmath,mathrsfs}
\usepackage{amssymb, amsmath}
\usepackage{graphicx}

\def\inte#1{
\displaystyle\mathop{#1\kern0pt}^\circ }

\def\virgp{\raise 2pt\hbox{,}}
\def\cdotpv{\raise 2pt\hbox{;}}

\def\C{\mathop{\mathbb C\kern 0pt}\nolimits}
\def\DD{\mathop{\mathbb D\kern 0pt}\nolimits}
\def\EE{\mathop{{\mathbb E \kern 0pt}}\nolimits}
\def\K{\mathop{\mathbb K\kern 0pt}\nolimits}
\def\N{\mathop{\mathbb N\kern 0pt}\nolimits}
\def\Q{\mathop{\mathbb Q\kern 0pt}\nolimits}
\def\R{\mathop{\mathbb R\kern 0pt}\nolimits}
\def\SS{\mathop{\mathbb S\kern 0pt}\nolimits}
\def\ZZ{\mathop{\mathbb Z\kern 0pt}\nolimits}
\def\TT{\mathop{\mathbb T\kern 0pt}\nolimits}
\def\P{\mathop{\mathbb P\kern 0pt}\nolimits}

\newcommand{\beq}{\begin{equation}}
\newcommand{\eeq}{\end{equation}}
\newcommand{\ben}{\begin{eqnarray}}
\newcommand{\een}{\end{eqnarray}}
\newcommand{\beno}{\begin{eqnarray*}}
\newcommand{\eeno}{\end{eqnarray*}}

\renewcommand{\theequation}{\thesection.\arabic{equation}}

\begin{document}

\title[LSTM for Volatility Prediction with Baidu Index]
{Long Short-Term Memory Networks for CSI300 Volatility Prediction with Baidu Search Volume}

\author[Y. -L. Zhou, R. -J. Han, Q. Xu and W. -K. Zhang]{Yu-Long Zhou,  School of Mathematics, Yunnan Normal University, Kunming,  P. R.  China. \\
Ren-Jie Han$^*$, College of Economics, Sichuan University, Chengdu, P. R. China \\
Qian Xu, College of Economics, Sichuan University, Chengdu, P. R. China \\
Wei-Ke Zhang, College of Economics, Sichuan University, Chengdu, P. R. China}

\thanks{$^*$ Corresponding author: Ren-Jie Han, 512910603@qq.com}

\begin{abstract}  Intense volatility in financial markets affect humans worldwide. Therefore, relatively accurate prediction of volatility is critical. We suggest that massive data sources resulting from human interaction with the Internet may offer a new perspective on the behavior of market participants in periods of large market movements. First we select 28 key words, which are related to finance as indicators of the public mood and macroeconomic factors. Then those 28 words of the daily search volume based on Baidu index are collected manually, from June 1, 2006 to October 29, 2017. We apply a Long Short-Term Memory neural network to forecast CSI300 volatility using those search volume data.  Compared to the benchmark GARCH model, our forecast is more accurate, which demonstrates the effectiveness of the LSTM neural network in volatility forecasting.
\end{abstract}

\maketitle

\tableofcontents




\renewcommand{\theequation}{\thesection.\arabic{equation}}
\setcounter{equation}{0}

\section{Introduction}
Volatility prediction in financial markets is of great practical and theoretical interest. Volatility plays crucial roles in financial markets, such as in derivative pricing, portfolio risk management, and hedging strategies. Therefore, it is demanding to find a ¡°good¡± method to more accurately forecast volatility.

According to Herbert Simon, actors begin their decision-making process by attempting to gather information \cite{Simon}. Nowadays, information gathering often consists of searching online sources. Hence search volumes of key words related to finance may reveal market sense and and focus of investors.  Similar to the Google Trends, the Baidu Index, based on Baidu Search, shows aggregated information on the volume of queries for different search terms and how these volumes change over time. Unlike Google trends, Baidu Index's search volume data can not be downloaded. The 28 key words we use in this article all come from manual collection.

In this study, we investigate the intriguing possibility of analyzing search query data from Baidu Index, modeled by Long Short-Term Memory neural network,  to show the feasibility of predicting the stock market volatility through the search volumes. To our best knowledge, there has been no previous attempt to deploy LSTM networks on a large and liquid Baidu Index's searching volume to assess its performance in stock market volatility prediction tasks. Besides, we compare the results of the LSTM against GARCH.

The increasing volumes of 'big data' reflecting various aspects of our everyday activities represent a vital new opportunity for scientists to address fundamental questions about the complex world we inhabit \cite{King,LaPeAdArBaBr,PeTeHaStPe}. Baidu index and Google search volumes, do not only reflect aspects of the current state of the economy, but may also provide some insight into future trends in the behavior of economic participants. Yu, Zhang (2012), taking Baidu search terms as an agent variable of personal investor concern, found that Baidu search volume can cause positive price pressure on the market in the current period, and this pressure will quickly reverse in next periods. Meanwhile, they found that investors' attention on non-trading days is significantly related to the price jump of stocks on the next trading day \cite{YuZh}. Baidu search volume is also used as a proxy variable for individual investors' attention in Zhao, Lu and Wang(2013). The authors analyzed the relationship between Baidu search volumes and the stock returns of 1301 stocks in Growth Enterprises Market Board, and found that there is a positive correlation between search volume and stock returns \cite{ZhLuWa}. Da et.al(2011), proposed a measure of investor attention using search frequency in Google (Search Volume Index(SVI)), provided evidence that SVI captures the attention of retail investors, and found the relation between investor attention and asset prices \cite{ZhEnGa}. Using historic data from the period between Jan 2004 and Feb 2011, T Preis et.al(2013), found that detectable increases in Google search volumes for keywords relating to financial markets before stock market falls \cite{PrMoSt}.

Prediction tasks on financial time series are notoriously difficult, primarily driven by the high degree of noise and the generally accepted, semi-strong form of market efficiency \cite{Fama}. Meanwhile, there are plenty of well-known capital market anomalies that are in stark contrast with the notion of market efficiency. In the past years, initial evidence has been established that machine learning techniques are capable of identifying (non-linear) structures in financial market data \cite{Huck,MoZi,DiKlJi}. Petersen, A et.al(2017) applied LSTM networks to all S\&P 500 constituents from 1992 until 2015 \cite{FiKr}.

\section{Data description and preprocessing}
In this work, we study the CSI 300 index based on publicly available daily data comprising high, low, open, close, and close prices. Daily returns $r_{t}$ are evaluated as the log difference of the close price, while daily volatility $h_{t}$ is estimated using the high, low, open and close prices in the following equation \cite{GaKl}.
\ben u_{t}=\log \frac{Hi_{t}}{Op_{t}}, d_{t}=\log \frac{Lo_{t}}{Op_{t}}, c_{t}=\log \frac{Cl_{t}}{Op_{t}}.\een
\ben h_{t}=0.511(u_{t}-d_{t})^{2}-0.019[c_{t}(u_{t}+d_{t})-2u_{t}d_{t}]-0.383c_{t}^{2}.\een
We remark that this definition is the best among all quadratic combination under some criteria \cite{GaKl}.

Starting from June 1st 2006, Baidu has been collecting the daily volume of searches from personal computer
related to various aspects of macroeconomics. This database is available to the public as the Baidu index PC trend $d_{t}$. Unfortunately, the Baidu index can not be downloaded.
\cite{YuZh} and \cite{ZhLuWa} have shown correlations between Baidu index and the equity market. In this work, we use this trend data as a representation of the public interest in various macroeconomic
factors.

For this study, we include 28 domestic trends which are
listed in Table 1 with their abbreviations.

\begin{table}[!htbp]
\centering
\caption{Search terms and their abbreviations}\label{terms}
\begin{tabular}{cc}
\hline
Exact search terms& Abbreviation\\
\hline
insurance& insur\\
fiscal revenue& fisrev\\
loan& loan\\
anti-corruption& anti-cor\\
real estate& reaest\\
debt& debt\\
leverage& lever\\
equity& equity\\
advertisement& adver\\
air ticket& airtic\\
education& educa\\
marriage& marri\\
financial investment& fininv\\
financial derivatives& finder\\
economics& econo\\
profit& profi\\
travel& trave\\
auto buyers& autbuy\\
auto finance& autfin\\
luxury goods& luxgoo\\
inflation& infla\\
crisis& crisi\\
default& defau\\
office building& offbui\\
credit card& crecar\\
bank& bank\\
increase& incre\\
bond& bond\\
\hline
\end{tabular}
\end{table}

We use $X$ to denote the aggregated data,
\beno X=(X^{1},X^{2},\cdots,X^{30})=(r, h, d^{advert}, \cdots, d^{travel}).\eeno
We split the whole data set into a training set (80\%) and a test set (20\%). The training set ranges from 1-June-2006 to 17-July-2015 while the test set ranges from 20-July-2015 to 27-Oct-2017. Additionally, it is worth noting here that all these 30 time series are stationary in the sense that their unit-root null hypotheses have p-values less than 0.05 in the Augmented Dickey-Fuller test \cite{Fama}.

Preprocessing the time series with different observation interval and normalization window may cause corresponding difference of causality between the input and output.  Let$\triangle t$ be the observation interval
\ben \label{r-delta-t} r^{\triangle t}_{i} = \sum_{t = (i-1)\triangle t + 1}^{i\triangle t} r_{t}.\een
\ben \label{r-index-t} d^{\triangle t}_{i} = \frac{1}{\triangle t}\sum_{t = (i-1)\triangle t + 1}^{i\triangle t} d_{t}.\een
\ben \label{r-sigma-t} h^{\triangle t}_{i} = \sqrt{\sum_{t = (i-1)\triangle t + 1}^{i\triangle t} h^{2}_{t}}.\een
In this study, we aim to predict volatility, so we denote the next period volatility by $Y^{\triangle t}$,
\beno Y^{\triangle t}_{i}=h^{\triangle t}_{i+1}.\eeno
We use moving average values to normalize the above observed data.  With a look-back window $k$, a time series $Z$ is normalized to $Z^{k}$ defined by
\beno Z^{k}_{i}=\frac{Z_{i}-mean(Z_{i-k : i})}{std(Z_{i-k : i})}.\eeno
Each combination of $\triangle t$ and $k$ should determine an observation and normalization scheme with its unique predictive
power. We denote these schemes as $(\triangle t, k)$ and the resulting data as $(X^{\triangle t, k}, Y^{\triangle t, k})$.
In principle, one may apply learning models on each scheme and evaluate the accuracy of prediction on a validation set such that
the optimal scheme can be chosen. Alternatively, an information metric can be set up to select the optimal scheme which maximizes this metric. In this work, we use the mutual information between $X^{\triangle t, k}$ and $Y^{\triangle t, k}$ for each pair $(\triangle t, k)$.

Let us make a brief introduction to mutual information.
For any discrete random variable pair $(X,Y)$, let $p_{X,Y}$ be the joint probability function of $(X,Y)$. Let $p_{X}$ and $p_{Y}$ be the marginal probability function of $X$ and $Y$ respectively. The mutual information between $X$ and $Y$ is defined as
\ben \label{mutual-information}\mathcal{MI}(X,Y)=\sum_{x,y}p_{X,Y}(x,y)\log\frac{p_{X,Y}(x,y)}{p(x)p(y)}.\een
Assuming conditional independence between the input variables in $X^{\triangle t, k}$, the mutual information
can be broken down into a sum of the individual
components of $X^{\triangle t, k}$ with $Y^{\triangle t, k}$.
Therefore we choose $(\triangle t, k)$ to maximize
\ben \mathcal{MI}(X^{\triangle t, k},Y^{\triangle t, k}):=\sum_{i=1}^{n}\mathcal{MI}(X^{i,\triangle t, k}, Y^{\triangle t, k}).\een
One may try to use the time series $\{X^{i,\triangle t, k}_{j}, Y^{\triangle t, k}_{j}\}_{1\leq j\leq T}$ to empirically compute $\mathcal{MI}(X^{i,\triangle t, k}, Y^{\triangle t, k})$ according to (\ref{mutual-information}). However, note that the values in $\{X^{i,\triangle t, k}_{j}, Y^{\triangle t, k}_{j}\}_{1\leq j\leq T}$ should be unique due the accuracy of real data. More precisely, one has
\beno X^{i,\triangle t, k}_{j}\neq X^{i,\triangle t, k}_{l}, Y^{\triangle t, k}_{j}\neq Y^{\triangle t, k}_{l},~~~~ \text{if}~~j\neq l.\eeno
Then a direct calculation gives
\beno \mathcal{MI}(X^{\triangle t, k},Y^{\triangle t, k})=n\log T,\eeno
which only depends on the sample size. Thus applying mutual information in this way can not reveal any relation between $X^{\triangle t, k}$ and $Y^{\triangle t, k}$. Considering this, we divide the data into small groups and regard values in each group as one point. Precisely, if $(X_{j}, Y_{j})_{1 \leq j \leq T}$ is a sample with size $T$, set $m_{X}=\min_{j}\{X_{j}\}, M_{X}=\max_{j}\{X_{j}\}, m_{Y}=\min_{j}\{Y_{j}\}, M_{Y}=\max_{j}\{Y_{j}\}$. Take a positive integer $N$, which is interpreted as the group number. Now divide the interval $[m_{X}, M_{X}]$ evenly into $N$ subintervals $\{I_{i}\}_{1\leq i\leq N}$.
Then set
\beno I_{i} = [m_{X}+(i-1)\frac{M_{X}-m_{X}}{N},m_{X}+(i-1)\frac{M_{X}-m_{X}}{N}[,~~1\leq i\leq N-1; I_{N} = [\frac{m_{X}}{N}+\frac{N-1}{N}M_{X}].\eeno
Similarly, divide the interval $[m_{Y}, M_{Y}]$ evenly into $N$ subintervals $\{J_{i}\}_{1\leq i\leq N}$. Then we define the marginal law function as
\beno P_{X}(i)=\frac{\text{card}\{m:X_{m}\in I_{i}\}}{T},~~P_{X}(j)=\frac{\text{card}\{l:Y_{l}\in I_{j}\}}{T}, ~~1\leq i, j\leq N,\eeno
and the joint law function as
\beno P_{X,Y}(i,j)=\frac{\text{card}\{(m,l):X_{m}\in I_{i},Y_{l}\in I_{j}\}}{T},  ~~1\leq i, j\leq N.\eeno
Then we define the empirical mutual information between series $(X_{j})_{1 \leq j \leq T}$  and $(Y_{j})_{1 \leq j \leq T}$  as
\ben \label{N-mutual-information}\mathcal{MI}^{N}(X,Y)=\sum_{i,j}p_{X,Y}(i,j)\log\frac{p_{X,Y}(i,j)}{p_{X}(i)p_{Y}(j)}.\een
Note that
\beno \lim_{N \rightarrow \infty}\mathcal{MI}^{N}(X,Y)=\log T,\eeno
which indicates we should not take too large $N$. In our study, we take $N=100$.

\begin{figure}[htbp]
\large
\centering
\includegraphics[scale=0.6]{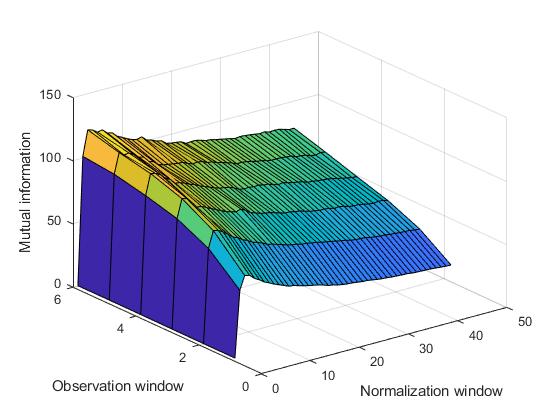}
\caption{Comparison of different observation and normalization window}
\label{3dmuin}
\end{figure}

Figure \ref{3dmuin} shows the mutual information for different combination of $(\triangle t, k)$. Clearly, when the normalization window $k$ is around 5, the mutual information is maximized.
Another obvious phenomenon is that, the longer the observation interval $\triangle t$, the larger the mutual information. However, due to the limited sample size, we can not choose too large $\triangle t$. To retain sufficient sample data, we select
\beno \triangle t =5, ~~k=5.\eeno
Note that $\triangle t =5$ corresponds to weekly return and volatility, which is also a consideration of our choice.

\section{Methods}
LSTM networks, introduced by Hochreiter and Schmidhuber (1997) and were furthered in the following years by Gers et al. (2000) and Graves and Schmidhuber (2005), belong to the class of recurrent neural networks (RNNs). LSTM networks are designed to learn long-term dependencies and are capable of vanishing and exploding gradients \cite{HaAnFr}.

\begin{figure}[htbp]
\large
\centering
\includegraphics[scale=1.0]{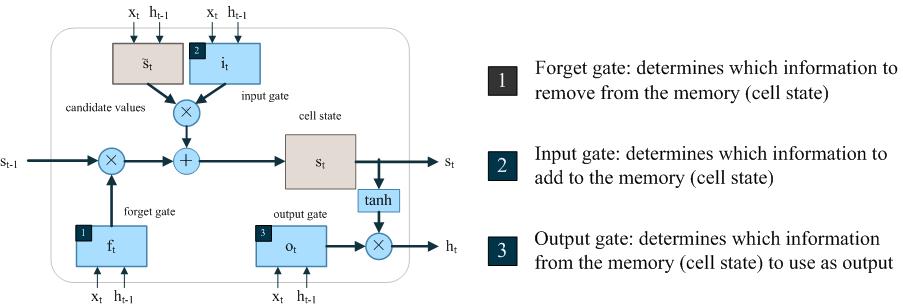}
\caption{Structure of LSTM memory cell}
\label{lstm}
\end{figure}

LSTM networks contain an input layer, one or more hidden layers, and an output layer.  The number of explanatory variables (feature space) equal to the number of neurons in the input layer. The output space is determined by the quantity of neurons in the output layer.
The hidden layer(s) contains the memory cell, which is the unique part of LSTM networks. Through three gates in each of the memory cells, the network maintains and adjusts its cell state $s_{t}$: an input gate $i_{t}$, a forget gate  $f_{t}$, and an output gate $o_{t}$. Figure \ref{lstm} shows the structure of a memory cell.

In figure \ref{lstm}, $x_{t}$ stands for the input vector at time step t, the information flow $s_{t-1}$  and the volatility estimation  $h_{t-1}$ are computed from the former step. $W_{f}, W_{s}, W_{i}$ and $W_{o}$ are weight matrices, $b_{f}, b_{s}, b_{i}$ and $b_{o}$ are bias vectors;  $f_{t}, i_{t}, o_{t}$ represent the values of each gate;  $s_{t}$ and $\tilde{s}_{t}$ are the cell gate and candidate values; $h_{t}$ is the output of vector at time t.

At step 1, the previous cell state $s_{t-1}$ is determined by the LSTM layer how much it should be forgotten. Given $x_{t}, h_{t-1}$, the bias term of the forget gate, $f_{t}$ can be computed as
\ben\label{forget-fraction} f_{t} = sigmoid((x_{t}, h_{t-1})W_{f}+b_{f}),\een
where the function $sigmoid$ is defined by
\beno sigmoid(x)=\frac{1}{1+\exp(-x)}.\eeno

 At step 2, the LSTM layer determines which information should be added to the network's cell states:

 \ben\label{tilde-s} \tilde{s}_{t} = \tanh((x_{t}, h_{t-1})W_{\tilde{s}}+b_{\tilde{s}}).\een

\ben\label{input} i_{t} = sigmoid((x_{t}, h_{t-1})W_{i}+b_{i}).\een
Here $\tanh(x)=\frac{\exp(x)-\exp(-x)}{\exp(x)+\exp(-x)}$.

 In the last step, the output $h_{t}$  is computed  through the following two equations:
\ben\label{output-gate} o_{t} = sigmoid((x_{t}, h_{t-1})W_{o}+b_{o}).\een
\ben\label{output-final} h_{t} = o_{t}\tanh(s_{t}).\een

Through (\ref{output-gate}) and (\ref{output-final}) the volatility $h_{t}$ will be predicted. The fundamental of this time series forecasting is
\ben\label{learn-forget} s_{t} = f_{t}s_{t-1}+i_{t}\tilde{s}_{t}.\een
We apply the deep learning library Keras to estimate the coefficients by training in python 3. Specifically, the lag of the LSTM is set at 50, and the bach contains 5 examples, time step is 5. The objective loss function we choose in the model is mean absolute percent error (MAPE). When we set the epochs at 200, the MAPE of the test set will drop to the minimum. We use 20\% of the observed data as the test set. Moreover, in order to evaluate the performance of the LSTM, we apply one autoregressive model (GARCH) as benchmark model.
\ben\label{garch} h^{2}_{t}=\omega+h^{2}_{t-1}(\alpha+\beta \epsilon^{2}_{t}), ~~\epsilon \sim \mathcal{N}(0,1).\een

\section{Results and discussion}
\begin{figure}[htbp]
\large
\centering
\includegraphics[scale=0.60]{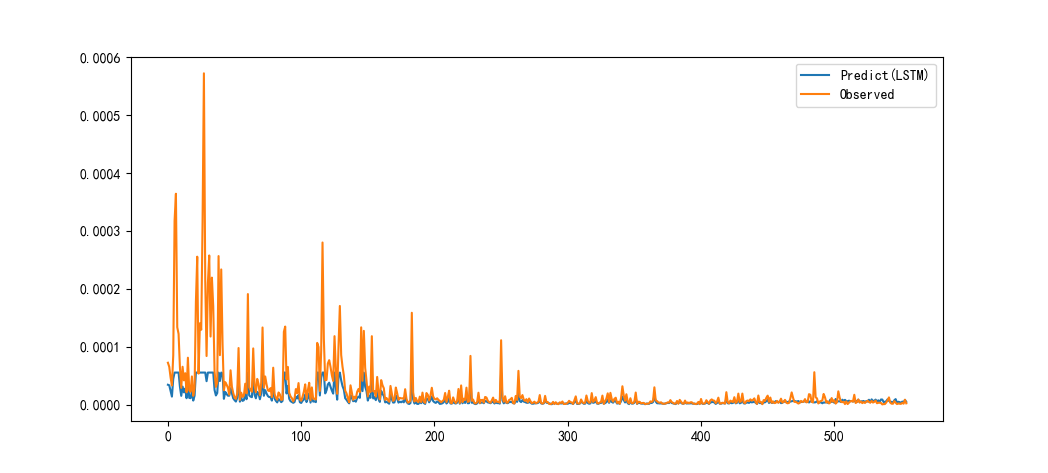}
\caption{Predicted volatility VS actual volatility}
\label{preVact}
\end{figure}

In figure \ref{preVact}, the observed volatility together with the predicted values is plotted. As we can see in the figure, the predicted values fit the actual volatility in decent accuracy, especially when the actual volatility is small.

As we have indicated in the last chapter, MAPE as the loss function, is shown in table \ref{errorComparison}. In terms of mean square error (MSE), the LSTM also performs better than the benchmark model.

\begin{table}[!htbp]
\centering
\caption{Comparison between LSTM and GARCH}\label{errorComparison}
\begin{tabular}{ccc}
\hline
Model& MSE & MAPE\\
\hline
LSTM&0.02& 0.17\\
GARCH&0.39& 1.25\\
\hline
\end{tabular}
\end{table}

Our LSTM model avoids significant over-fitting as the MAPE evaluated in the test set is 17\%, which is close to the MAPE (\%15.6) in the training set. The MSE of the LSTM model is 2\% which is far smaller than the benchmark model, as listed in table \ref{errorComparison}.

We chose the intermediate timescale with hand collecting Baidu Index's search volume to investigate the performance of the LSTM model on stock market volatility prediction.
If we have high-frequency Baidu Index's searching volume, we may also forecast high-frequency volatility, which is much more interesting and practical. The input of the neural network could be replaced by the market micro-structure information \cite{AbBoFoLeRo}. By maximizing the mutual information, we can find suitable input feature set, normalization window and observation interval.

When we want to expand the usage scenario of the model, such as studying the volatility of individual stock or the volatility in different financial market, we could change the input or the structure of the LSTM layer. In this paper, we take the Baidu Index's searching volume as the hidden market state, and discuss the feasibility of using it to estimate the volatility of CSI 300. The LSTM model is far more efficient than the benchmark model. We have investigated the autocorrelation and partial autocorrelation functions of the error series in the test set, which shows that the prediction error has no memory of itself.

\section{Conclusion}
In this work, we have successfully demonstrated that the LSTM network is able to effectively extract meaningful information from noisy financial time series data. We consider the Baidu Index's searching volume as proxy variables. Together with (or as) the market information, they shows the power of reflecting the CSI300 daily volatility change. We apply one layer LSTM neural network and select 80\% of the whole data set as the training set. In terms of MAPE and MSE, the LSTM model outperforms the benchmark model. In addition, we collect 28 key words of the daily search volume of the PC end. Considering the development of mobile technology and the increase of mobile device search, it is not appropriate to estimate the CSI 300 volatility only using the PC end search. This is also the main limitation of our paper. In the future research, we will collect searching volumes from mobile device to perfect our forecast.

 {\bf Acknowledgments.} Any reader who is interested in our study may contact the corresponding author to get the data we used.

\end{document}